\documentclass[aps,pre,twocolumn,groupedaddress]{revtex4-1}

\usepackage{xcolor}

\usepackage{amsfonts,amssymb,amsmath} 
\usepackage{color} 
\usepackage{graphicx} 
\usepackage{epstopdf,mathrsfs} 
\usepackage[colorlinks,linktocpage,linkcolor=blue]{hyperref}

\begin{document}


\title{L\'{e}vy-walk-like Langevin dynamics affected by a time-dependent force}

\author{Yao Chen}
\author{Weihua Deng}

\affiliation{School of Mathematics and Statistics, Gansu Key Laboratory
of Applied Mathematics and Complex Systems, Lanzhou University, Lanzhou 730000,
P.R. China}

\begin{abstract}

L\'{e}vy walk is a popular and more `physical' model to describe the phenomena of superdiffusion, because of its finite velocity. The movements of particles are under the influences of external potentials almost at anytime and anywhere. In this paper, we establish a Langevin system coupled with a subordinator to describe the L\'{e}vy walk in the time-dependent periodic force field. The effects of external force are detected and carefully analyzed, including nonzero first moment (even though the force is periodic), adding an additional dispersion on the particle position, the consistent influence on the ensemble- and time-averaged mean-squared displacement, etc. Besides, the generalized Klein-Kramers equation is obtained, not only for the time-dependent force but also for space-dependent one.


\end{abstract}


\maketitle

\section{Introduction}

{The universal existence of the anomalous diffusion phenomena in the natural world has stimulated the exploration and research of scientists, in which the ensemble-averaged mean-squared displacement (EAMSD)
\begin{equation}
  \langle(\Delta x(t))^2\rangle:=\langle[x(t)-\langle x(t)\rangle]^2\rangle\propto t^\alpha
\end{equation}
is the commonly used statistical quality for describing diffusion phenomena \cite{JeonMetzler:2012,EuleFriedrich:2009,CairoliBaule:2015,MagdziarzWeronKlafter:2008,ChenWangDeng:2017}.} Except EAMSD, another important statistic to study particle diffusion property is the time-averaged mean-squared displacement (TAMSD), which is defined as  \cite{MetzlerJeonCherstvyBarkai:2014,BurovJeonMetzlerBarkai:2011,AkimotoCherstvyMetzler:2018}
\begin{equation}\label{TADef}
\begin{split}
 \overline{\delta^2(\Delta)}
  &=\frac{1}{T-\Delta}\int_0^{T-\Delta}  [x(t+\Delta)-x(t)  \\
  &~~~    -\langle x(t+\Delta)-x(t)\rangle]^2dt.
\end{split}
\end{equation}
The TAMSD can be obtained by analyzing the time series of single trajectory of particle in experiments. Here $\Delta$ is the lag time, $T$  the measurement time. In order to catch the statistical properties, the total measurement time $T$ needs to be much longer than the lag time $\Delta$. The single particle tracking techniques have been widely applied in studying the particles diffusion in living cell \cite{GoldingCox:2006,WeberSpakowitzTheriot:2010,BronsteinIsraelKeptenMaiTalBarkaiGarini:2009}.
For an unbiased particle, the mean value of particle's displacement in the EAMSD or TAMSD disappears. The equivalence of EAMSD and TAMSD as the measurement time $T\rightarrow\infty$ indicates the ergodicity of the stochastic process, such as Brownian motion.

The anomalous diffusion process, characterized by the EAMSD with power-law index $\alpha\neq1$, can be described by many kinds of physical models, the most popular of which is L\'{e}vy walk \cite{ZaburdaevDenisovKlafter:2015,SanchoLacastaLindenbergSokolovRomero:2004,RebenshtokDenisovHanggiBarkai:2014,ZaburdaevDenisovHanggi:2013}. In L\'{e}vy walk, the particle moves on a straight line for a random time. During this time period, the velocity of the particle maintains a fixed value. Then at the end of the excursion, the particle will choose a new direction randomly and move for another random time. The value of the velocity in this time period is as same as the last excursion. L\'{e}vy walk model seems to be more reasonable since it can characterize particle's motion with finite velocity.  This model is originally characterized by a coupled continuous time random walk (CTRW) \cite{ShlesingerKlafterWong:1982,KlafterBlumenShlesinger:1987,Zaburdaev:2006,ZaburdaevDenisovKlafter:2015}, in which the probability density functions (PDFs) of jump lengths and flight times are coupled by a constant velocity. According to the power-law exponent $\alpha$ of the PDF of flight time, L\'{e}vy walk expresses ballistic diffusion ($0<\alpha<1$), sub-ballistic superdiffusion ($1<\alpha<2$), and normal diffusion ($\alpha>2$) behavior \cite{ZaburdaevDenisovKlafter:2015}. Due to the finite velocity and multiple types of diffusions expressed by this model, it has been widely applied, not only in the tracking studies of animals or humans \cite{Nathan.etal:2008}, but also the anomalous superdiffusion of cold atoms in optical lattices \cite{KesslerBarkai:2012}, endosomal active transport within living cells \cite{ChenWangGranick:2015}, etc.

In addition to the diffusion behavior and the shape of PDFs, the ergodic behavior of the free L\'{e}vy walk has also been investigated \cite{FroembergBarkai:2013,GodecMetzler:2013}. When
the power-law exponent satisfies $1<\alpha<2$ resulting in a finite moment of flight time, the ensemble-averaged TAMSD and EAMSD only differ by a constant factor for a long measurement time. This phenomenon is named as an `ultraweak' non-ergodic behavior \cite{GodecMetzler:2013}. Similar to the case of $\alpha\in(1,2)$, the ensemble-averaged TAMSD also differs from EAMSD by a constant factor when $\alpha\in(0,1)$. But, because of the divergent first moment of flight times for $\alpha\in(0,1)$, the TAMSD is not self-averaged even when the measurement time $T\rightarrow \infty$ \cite{FroembergBarkai:2013}.

In real life, a particle seldom moves in a completely free environment. Most of the time, it is under the influence of an external force, which may depend on space or time.
In the case of harmonic potential, an overdamped Langevin equation was established \cite{WangChenDeng:2020PRE}; and the EAMSD of confined L\'{e}vy-walk-like Langevin dynamics can be obtained directly from the velocity correlation function for the force-free case.
The EAMSD grows to a stationary value for any $\alpha$, while the TAMSD keeps growing for $0<\alpha<1$, which indicates the non-ergodic behavior of the confined L\'{e}vy walk. When $1<\alpha<2$, the TAMSD approaches twice the stationary value of EAMSD, similar to confined fractional Brownian motion and fractional Langevin equation \cite{JeonMetzler:2012}.
L\'{e}vy walk under an external constant force can be described by the collision model \cite{BarkaiFleurov:1998} or a Langevin system coupled with a subordinator \cite{ChenWangDeng:2019-3}; utilizing the four-point joint PDF of the inverse subordinator, one finds that under the influence of external constant force, the L\'{e}vy walk particles always show super-ballistic diffusion phenomenon. More specifically, the EAMSD behaves as $t^4$ and $t^{5-\alpha}$ when $0<\alpha<1$ and $1<\alpha<2$ respectively, which is different from the TAMSD behaving as $T^2\Delta^2$ when $0<\alpha<1$ and $T^{3-\alpha}\Delta^2$ when $1<\alpha<2$. The non-ergodicity of L\'{e}vy walk under a constant force is obvious. In addition, for the case of constant force, the generalized Einstein relation \cite{BouchaudGeorges:1990,BarkaiMetzlerKlafter:2000,MetzlerKlafter:2000} for the EAMSD is still satisfied while it does not hold for the TAMSD \cite{FroembergBarkai:2013-3}.

In this paper, we focus on how the external time-dependent periodic force affects the L\'{e}vy walk. The case of periodic force acting on the subdiffusive CTRW has been discussed in Refs. \cite{SokolovKlafter:2006,MagdziarzWeronKlafter:2008,ChenWangDeng:2019-2}. Here, we establish a set of Langevin equations coupled with a subordinator to describe the L\'{e}vy-walk-like Langevin dynamics with a time-dependent force. Based on the Langevin system and the two-point joint PDF of the inverse subordinator, the velocity correlation function, and further the EAMSD and the TAMSD can be obtained. We find that the first moment of the particle's displacement is not null, although the external force is periodic, which is different from the result of subdiffusive CTRW. Besides, the external periodic force brings an additional dispersion to this system without changing the diffusion behavior and retains the `ultraweak' non-ergodic behavior of the free L\'{e}vy walk. The corresponding generalized Klein-Kramers equation satisfied by the joint PDF $P(x,v,t)$ is also derived in this paper, not only for the  time-dependent force but also for a general space-dependent one.

The structure of this paper is as follows. In Sec. \ref{two_s}, we review the (inverse) subordinator, the relationship of the moments, and the correlation functions between the original and subordinated processes. In Sec. \ref{three}, we present the Langevin picture
of the L\'{e}vy walk affected by time-dependent force for all times. In Secs. \ref{four}-\ref{six}, the first moment, the velocity correlation function, and the MSDs are evaluated, respectively, to show the influence of external periodic force on the L\'{e}vy-walk-like Langevin dynamics. The corresponding generalized Klein-Kramers equation is derived in Sec. \ref{seven}. Finally, we make the summaries in Sec. \ref{eight}.

\section{Subordinator}\label{two_s}

Subordinator is a non-decreasing L\'{e}vy process with stationary and independent increments  \cite{Applebaum:2009}. Its characteristics determine that it can depict the evolution of time. To characterize the power-law distributed flight time of the L\'{e}vy walk, we take the $\alpha$-dependent subordinator $t(s)$, the characteristic function of which is $p(\lambda,s)=\langle{e}^{-\lambda t(s)}\rangle={e}^{-s\Phi(\lambda)}$. When $0<\alpha<1$, we have $\Phi(\lambda)=\lambda^\alpha$ \cite{BauleFriedrich:2005};  besides, when $1<\alpha<2$,  $\Phi(\lambda)= \tau_0/(\alpha-1)\lambda-{\tau_0^\alpha} |\Gamma(1-\alpha)| \lambda^\alpha$ \cite{WangChenDeng:2019}. The brackets $\langle\cdots\rangle$ denote the statistical average over many stochastic realizations.
The two-point PDF $p(t_1, s_1; t_2, s_2)$ in Laplace space $(t_1\rightarrow\lambda_1, t_2\rightarrow\lambda_2)$ can be obtained by use of the independence of the increments of subordinator $t(s)$  \cite{BauleFriedrich:2005}
\begin{equation}\label{pp}
\begin{split}
&p(\lambda_1, s_1; \lambda_2, s_2)\\
&=\langle e^{-\lambda_1 t(s_1)}e^{-\lambda_2 t(s_2)}\rangle\\
&=\Theta(s_2-s_1)e^{-s_1\Phi(\lambda_1+\lambda_2)}e^{-(s_2-s_1)\Phi(\lambda_2)}\\
&~~~+\Theta(s_1-s_2)e^{-s_2\Phi(\lambda_1+\lambda_2)}e^{-(s_1-s_2)\Phi(\lambda_1)}.
\end{split}
\end{equation}

The corresponding inverse process, named inverse $\alpha$-dependent subordinator $s(t)$, has the definition \cite{KumarVellaisamy:2015,AlrawashdehKellyMeerschaertScheffler:2017}
\begin{equation}
s(t)=\inf_{s>0}\{s:t(s)>t\},
\end{equation}
which can  be regarded as the first-passage time of the subordinator $\{t(s),\,s\geq 0\}$.
The PDF of the inverse $\alpha$-dependent subordinator $s(t)$, defined as $h(s,t)=\frac{\partial}{\partial s}\langle \Theta(s-s(t))\rangle$, has the Laplace transform ($t\rightarrow\lambda$) \cite{BauleFriedrich:2005}
\begin{equation}\label{hslambda}
h(s,\lambda)
=\frac{\Phi(\lambda)}{\lambda}e^{-s\Phi(\lambda)},
\end{equation}
which can be obtained through the relationship with the PDF $p(t,s)=\frac{\partial}{\partial t}\langle \Theta(t-t(s))\rangle$ of subordinator $t(s)$: $\langle \Theta(s-s(t))\rangle=1-\langle \Theta(t-t(s))\rangle$.
The two-point PDF $h(s_1, t_1; s_2, t_2)$ of the inverse subordinator $s(t)$ in Laplace space is \cite{BauleFriedrich:2005,WangChenDeng:2019}
\begin{equation}\label{hh}
\begin{split}
&h(s_1,\lambda_1;s_2,\lambda_2)  \\
    &=\frac{\partial}{\partial s_1} \frac{\partial}{\partial s_2} \frac{1}{\lambda_1\lambda_2}\,p(\lambda_1,s_1;\lambda_2,s_2) \\
    &=\delta(s_2-s_1)\frac{\Phi(\lambda_1)+\Phi(\lambda_2)-\Phi(\lambda_1+\lambda_2)}{\lambda_1\lambda_2}\,{e}^{-s_1\Phi(\lambda_1+\lambda_2)} \\
    &~~~+\Theta(s_2-s_1)\frac{\Phi(\lambda_2)(\Phi(\lambda_1+\lambda_2)-\Phi(\lambda_2))}{\lambda_1\lambda_2} \\
    &~~~~~~\times{ e}^{-s_1\Phi(\lambda_1+\lambda_2)}{e}^{-(s_2-s_1)\Phi(\lambda_2)}   \\
    &~~~+\Theta(s_1-s_2)\frac{\Phi(\lambda_1)(\Phi(\lambda_1+\lambda_2)-\Phi(\lambda_1))}{\lambda_1\lambda_2} \\
    &~~~~~~\times{ e}^{-s_2\Phi(\lambda_1+\lambda_2)}{e}^{-(s_1-s_2)\Phi(\lambda_1)},
\end{split}
\end{equation}
where the first equality comes from the relation
\begin{equation}
\begin{split}
& \langle \Theta(s_2-s(t_2))\Theta(s_1-s(t_1))\rangle
    = 1-\langle\Theta(t_2-t(s_2))\rangle  \\
   &~~~ -\langle\Theta(t_1-t(s_1))\rangle
 +\langle\Theta(t_2-t(s_2))\Theta(t_1-t(s_1))\rangle,
\end{split}
\end{equation}
and the second equality is gotten by use of the two-point PDF $p(t_1, s_1; t_2, s_2)$ in Eq. \eqref{pp}.


The PDFs of inverse subordinator in Eqs. \eqref{hslambda} and \eqref{hh} act as a bridge between the subordinated process and the original process. Let the original process be $y(s)$ with the PDF $g_0(y,s)$, and the subordinated process  $y(t):=y(s(t))$ with PDF $g(y,t)$. Then it holds that \cite{BauleFriedrich:2005,Barkai:2001,ChenWangDeng:2018-2}
\begin{equation}
g(y, t)=\int_0^\infty  g_0(y,s)h(s,t) ds.
\end{equation}
Further, the moments of the subordinated process $y(t)$ in Laplace space is
\begin{equation}\label{one}
\langle y^n(\lambda)\rangle=\int_0^\infty \langle y^n(s)\rangle h(s,\lambda) ds.
\end{equation}
Similarly, the two-point PDF $g(y_1, t_1;y_2,t_2)$ of the subordinated process $y(t)$ can be connected with the two-point PDF $g_0(y_1,s_1;y_2,s_2)$ of the original process $y(s)$ as
\begin{equation}
\begin{split}
&g(y_1, t_1;y_2,t_2)\\
&=\int_0^\infty \int_0^\infty g_0(y_1,s_1; y_2, s_2)h(s_1,t_1;s_2,t_2)ds_1ds_2.
\end{split}
\end{equation}
Then the correlation function of $y(t)$ in Laplace space is
\begin{equation}\label{two}
\begin{split}
&\langle y(\lambda_1)y(\lambda_2)\rangle\\
&=\int_0^\infty \int_0^\infty\langle y(s_1)y(s_2)\rangle h(s_1,\lambda_1;s_2,\lambda_2)ds_1 ds_2.
\end{split}
\end{equation}

In the rest of this paper, we establish the Langevin equation with an external force, coupled with an $\alpha$-dependent subordinator, to describe the L\'{e}vy walk in the external time-dependent force field. Then by use of the formulas \eqref{one} and \eqref{two}, we mainly evaluate some statistical qualities to show how a L\'{e}vy walk particle responds to the time-dependent force field.

\section{L\'{e}vy walk with time-dependent force}\label{three}
We have proposed the Langevin picture of the free L\'{e}vy walk dynamics in Ref. \cite{WangChenDeng:2019}. It is convenient to include an external force and calculate the velocity correlation function in a Langevin system, which is also the reason why we study the L\'{e}vy-walk-like Langevin dynamics.
In order to inherit the advantages of Langevin equation, the Langevin picture of L\'{e}vy walk under a time-dependent force is exhibited here, which is presented by a set of Langevin equations coupled with a subordinator
\begin{equation}\label{LW_force}
\begin{split}
    \frac{d}{d t}x(t)&=v(t),\\
    \frac{d}{d s}v(s)&=-\gamma v(s) +F(t(s)) \eta(s)+\xi(s),\\
    \frac{d}{d s}t(s)&= \eta(s).
\end{split}
\end{equation}
Here $\gamma$ represents the friction coefficient, $F(t(s))$ characterizes the time-dependent force, and $\xi(s)$ is a  Gaussian white noise. As we all know, the mean value of the Gaussian white noise is $\langle \xi(s)\rangle=0$ and the correlation function is $\langle \xi(s_1)\xi(s_2)\rangle=2D\delta(s_1-s_2)$. The L\'{e}vy noise $\eta(s)$ is considered as the formal derivative of the $\alpha$-dependent subordinator $t(s)$, which characterizes the distribution of each flight time of L\'{e}vy walk. The two noises, L\'{e}vy noise $\eta(s)$ and  Gaussian white noise $\xi(s)$, are independent.
The integral of velocity $v$ over physical time $t$ is the particle position $x$.
The initial position and initial velocity are both assumed to be null, i.e., $x(0)=v(0)=0$. Taking $F(t(s))=0$, the Langevin picture Eq. \eqref{LW_force} reduces to the force-free case \cite{WangChenDeng:2019}.

The key of Langevin system Eq. \eqref{LW_force} is its second equation. The variables $v(s)$ and $\xi(s)$ are given with respect to the operation time $s$, implying that the velocity of target particle is changed by the collision with surrounding small molecules along with the evolution of operation time $s$. By contrast, the external force in this equation is expressed as $F(t(s))$ (rather than $F(s)$) to indicate that it only makes sense as a function of physical time $t$ after the subordination in practice. The multiplier $\eta(s)$ balances the effect of external force made on physical time $t$ and the evolution of the equation in operation time $s$.

For further understanding of the external force term $F(t(s))\eta(s)$, we transform the second equation in Eq. \eqref{LW_force} to the one evolving over physical time $t$ by the technique of subordination (see Appendix \ref{App1} for the detailed derivation):
\begin{equation}\label{dvt}
\begin{split}
\frac{d}{d t}v(t)=-\gamma v(t) \frac{d}{d t}s(t)+F(t)+\xi(s(t))\frac{d}{d t}s(t).
\end{split}
\end{equation}
 It can be seen that the time-dependent force $F(t)$ acts on the system for the whole physical time. Especially for a trapping period when $s(t)$ is a constant, the direction of the motion of particle remains unchanged and the acceleration is exactly $F(t)$, i.e., $dv(t)/dt=F(t)$.

The closed form of velocity process in operation time $s$ is
\begin{equation}\label{vs}
\begin{split}
v(s)=&\int_0^s e^{-\gamma(s-s')}F(t(s'))\eta(s')ds'\\
&+\int_0^s e^{-\gamma(s-s')}\xi(s')ds',
\end{split}
\end{equation}
which is obtained by virtue of the Laplace transform method towards the second equation in Eq. \eqref{LW_force}. Then after the subordination, the expression of the velocity process in physical time $t$ is (see Appendix \ref{App1})
\begin{equation}\label{vt}
\begin{split}
v(t)&=\int_0^t e^{-\gamma(s(t)-s(t'))}F(t')dt'  \\
&~~~+\int_0^t e^{-\gamma(s(t)-s(t'))}\xi(s(t'))ds(t').
\end{split}
\end{equation}
The first term in the above expression comes from the external time-dependent force, and the second term from the random force $\xi$, which corresponds to the free L\'{e}vy walk.


In the following, mainly based on the velocity expression Eq. \eqref{vs} in operation time $s$, we firstly calculate the first moment and correlation function of velocity, and then evaluate the moments of particle displacement to study the effect of the time-dependent periodic force $F(t)=f_0 \sin(\omega t)$ on L\'{e}vy walk.

\section{First moment}\label{four}

It is well known that the first moment of the free stochastic process is null. Intuitively, one may expect that the first moment of the particle affected by a periodic force is also null.
To obtain the first moment of the L\'{e}vy walk under the time-dependent periodic force $F(t)=f_0 \sin(\omega t)$, we can rewrite the expression of $v(s)$ in Eq. \eqref{vs} as
\begin{equation}\label{vss}
\begin{split}
v(s)=&-\frac{f_0}{\omega}\int_0^s e^{-\gamma(s-s')}d\cos(\omega t(s'))\\
&+\int_0^s e^{-\gamma(s-s')}\xi(s')ds',
\end{split}
\end{equation}
where we have used the relationship $$F(t(s))\eta(s)=f_0 \sin(\omega t(s))\eta(s)=-\frac{f_0}{\omega}\frac{d}{ds}\cos(\omega t(s)).$$ Making the ensemble average toward this expression, one has
\begin{equation}
\begin{split}
\langle v(s)\rangle=-\frac{f_0}{\omega}\int_0^s e^{-\gamma(s-s')}d\langle \cos(\omega t(s'))\rangle\,
\end{split}
\end{equation}
with the mean of the noise $\xi$ vanishing. Using the fact that $\cos(\omega t)=\frac{1}{2}(e^{i\omega t}+e^{-i\omega t})$ and the characteristic function of subordinator $t(s)$: $\langle e^{\pm i\omega t(s)}\rangle=e^{- \Phi(\mp i\omega) s}$ for small $\omega$, one obtains the first moment of velocity process $v(s)$ as
\begin{equation}\label{v_s}
\begin{split}
\langle v(s)\rangle=&\frac{f_0}{2\omega}\frac{\Phi(-i\omega)}{\gamma-\Phi(-i\omega)}\left(e^{-\Phi(-i\omega)s}-e^{-\gamma s}\right)\\
                   &+\frac{f_0}{2\omega}\frac{\Phi(i\omega)}{\gamma-\Phi(i\omega)}\left(e^{-\Phi(i\omega)s}-e^{-\gamma s}\right).
\end{split}
\end{equation}
Through the formulas Eqs. \eqref{hslambda}, \eqref{one} and \eqref{v_s}, the first moment of velocity process $v(t)$ in Laplace space $(t\rightarrow \lambda)$ is
\begin{equation}
\begin{split}
&\langle v(\lambda)\rangle \\
&=
\frac{f_0}{2\omega}\frac{\Phi(i\omega)}{\gamma-\Phi(i\omega)}
\frac{\Phi(\lambda)}{\lambda}\left(\frac{1}{\Phi(i\omega)+\Phi(\lambda)}
-\frac{1}{\gamma+\Phi(\lambda)}\right) \\
&~~+\frac{f_0}{2\omega}\frac{\Phi(-i\omega)}{\gamma-\Phi(-i\omega)}
\frac{\Phi(\lambda)}{\lambda}\left(\frac{1}{\Phi(-i\omega)+\Phi(\lambda)}
-\frac{1}{\gamma+\Phi(\lambda)}\right) \\
&\simeq \frac{f_0}{\omega \gamma}\frac{\Phi(\lambda)}{\lambda},
\end{split}
\end{equation}
where we consider the asymptotics $\lambda\ll \omega$ in the last line.
After the inverse Laplace transform, one arrives at the first moment of velocity process for large time $t$, namely,
\begin{equation}\label{v-mean}
\begin{split}
\langle v(t)\rangle  \simeq \left\{
    \begin{array}{ll}
    \frac{f_0}{\omega \gamma\Gamma(1-\alpha)}t^{-\alpha}, &~~ 0<\alpha<1, \\[4pt]
    \frac{f_0}{\omega\gamma}(1+t/\tau_0)^{-\alpha}, &~~ 1<\alpha<2.
\end{array}
  \right.
\end{split}
\end{equation}
From the observation of the expression of $v(t)$ in Eq. \eqref{vt}, one can speculate the oscillation behavior of $\langle v(t)\rangle$ for short time since $e^{-\gamma(s(t)-s(t'))}\approx1$ for all $t'$ with small $t$. Then as time goes on, the linear response to external oscillation dies out, and $\langle v(t)\rangle$ tends to zero in Eq. \eqref{v-mean}. The tendency to zero is a quite different phenomenon from the CTRWs in a time-dependent periodic force \cite{SokolovKlafter:2006,ChenWangDeng:2019-2}. In CTRWs, the first moment tends to a positive constant when the force acts on the operation time $s$, while it still keeps oscillating for long time when the force acts on the physical time $t$. Here, the tendency to zero in Eq. \eqref{v-mean} might result from the fiction term $-\gamma v(s)$ in Eq. \eqref{LW_force}, which is the essential difference from the CTRWs in Refs. \cite{SokolovKlafter:2006,ChenWangDeng:2019-2}.

Since the mean velocity tends to zero, the speed to zero and the sign of velocity should be concerned. As Eq. \eqref{v-mean} shows, it tends to zero at the rate $t^{-\alpha}$ whatever $0<\alpha<1$ or $1<\alpha<2$, implying a faster decaying tendency for a larger $\alpha$.
In addition, the sign of velocity is consistent with the coefficient $f_0$, being positive. Since the linear response decays in course of the time, the bias to positive is yielded by the external perturbation at short times, where $\sin(\omega t)$ is positive.

\begin{figure}[!htb]
\begin{minipage}{0.5\linewidth}
  \centerline{\includegraphics[scale=0.5]{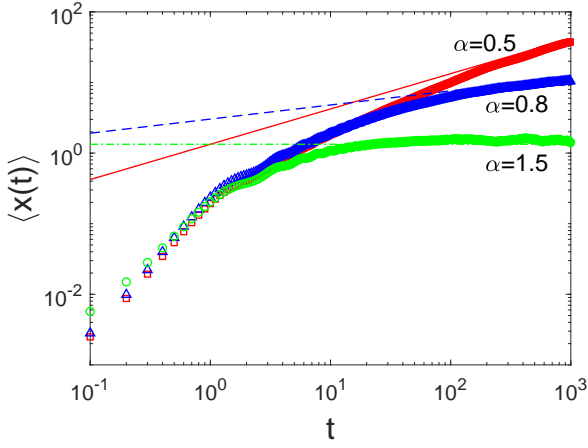}}
\end{minipage}
\caption{{(Color online)} First moment $\langle x(t)\rangle$ for the stochastic process described by Langevin equation Eq. \eqref{LW_force} with different $\alpha$. Other parameters are $f_0=1$, $\omega=3$, $\gamma=0.5$, and $\tau_0=1$.
From top to bottom, the red square-markers,
blue triangle-markers, and green circle-markers are the simulation results for $\alpha=0.5,~0.8,~1.5$, respectively. The red solid line,
blue dashed line, and green dot-dashed line are the corresponding theoretical results present in Eq. \eqref{x-mean}. }\label{fig1}
\end{figure}

Although the velocity tends to zero at the rate $t^{-\alpha}$, the mean value of displacement $x(t)$ behaves much different for different $\alpha$.
The integration of $\langle v(t)\rangle$ in Eq. \eqref{v-mean} leads to
\begin{equation}\label{x-mean}
\begin{split}
\langle x(t)\rangle  \simeq \left\{
    \begin{array}{ll}
    \frac{f_0}{\omega \gamma\Gamma(2-\alpha)}t^{1-\alpha}, &~~ 0<\alpha<1, \\[4pt]
    \frac{f_0\tau_0}{\omega\gamma(\alpha-1)}, &~~ 1<\alpha<2.
\end{array}
  \right.
\end{split}
\end{equation}
The time-dependent periodic force yields a nonzero mean value of the displacement, which is different from the zero mean in the case of a space-symmetric potential, such as harmonic potential. Here, $\langle x(t)\rangle$ grows for $0<\alpha<1$, but converges to a saturation value for $1<\alpha<2$ at a power-law rate $t^{1-\alpha}$. The simulation results are shown in Fig. \ref{fig1}, which agree with the theoretical result in Eq. \eqref{x-mean} for long time.

\section{velocity correlation function}\label{five}
To get the EAMSD and TAMSD, we need firstly obtain the velocity correlation function of the concerned process.
Due to the independence of the two terms in Eq. \eqref{vss}, the velocity correlation function in operation time consists of two parts $\langle v(s_1)v(s_2)\rangle=\langle v(s_1)v(s_2)\rangle_1+\langle v(s_1)v(s_2)\rangle_2$, where
\begin{equation}\label{VCF1}
\begin{split}
\langle v(s_1)v(s_2)\rangle_1=&\frac{f_0^2}{\omega^2}\int_0^{s_1} \int_0^{s_2} e^{-\gamma(s_1-s'_1)}e^{-\gamma(s_2-s'_2)}\\&\times\frac{d^2}
{ds'_1ds'_2}\langle \cos(\omega t(s'_1))\cos(\omega(t(s'_2)))\rangle
\end{split}
\end{equation}
and
\begin{equation}
\begin{split}
\langle v(s_1)v(s_2)\rangle_2=&\int_0^{s_1} \int_0^{s_2} e^{-\gamma(s_1-s'_1)}e^{-\gamma(s_2-s'_2)} \\&\times \langle \xi(s_1')\xi(s_2')\rangle ds_1'ds_2'.
\end{split}
\end{equation}
The second part can be easily obtained as $\langle v(s_1)v(s_2)\rangle_2= \frac{D}{\gamma} (e^{-\gamma |s_1- s_2|}-e^{-\gamma(s_1+s_2)})$. After using the subordination method Eq. \eqref{two}
and taking the inverse Laplace transform, for large $t_1$ and $t_2$ ($t_1<t_2$), one has
\begin{equation}\label{1}
\begin{split}
&\langle v(t_1)v(t_2)\rangle_2\\
&\simeq \left\{
\begin{array}{ll}
  \frac{D}{\gamma}\frac{\sin(\pi\alpha)}{\pi}B\left(\frac{t_1}{t_2};\alpha,1-\alpha \right),  & ~0<\alpha<1,  \\[5pt]
  \frac{D}{\gamma}\tau_0^{\alpha-1}\left((t_2-t_1)^{1-\alpha}-t_2^{1-\alpha}\right), &~ 1<\alpha<2,
\end{array}\right.
\end{split}
\end{equation}
where $B(x; a, b)$ is the incomplete Beta function. Equation \eqref{1} is as same as the velocity correlation function of free L\'{e}vy walk.

In the following, we pay attention to the first part $\langle v(s_1)v(s_2)\rangle_1$, which embodies the contribution of the time-dependent external force.
Using $\cos(\omega t)=\frac{1}{2}(e^{i\omega t}+e^{-i\omega t})$ again, one has
\begin{equation}\label{coss}
\begin{split}
&\langle \cos(\omega t(s_1))\cos(\omega t(s_2))\rangle\\
&=\frac{1}{4}[p(-i\omega,s_1;-i\omega,s_2)+p(-i\omega,s_1;i\omega,s_2)\\
&~~~+p(i\omega,s_1;-i\omega,s_2)+p(i\omega,s_1;i\omega,s_2)],
\end{split}
\end{equation}
where $p(\cdot)$ is the two-point PDF of the subordinator $t(s)$ defined in Eq. \eqref{pp} in Sec. \ref{two_s}. Taking the partial derivatives with respect to $s_1$ and $s_2$ on Eq. \eqref{coss} , the left-hand side connects to Eq. \eqref{VCF1} while the right-hand side can be expressed through the two-point joint PDF of the inverse subordinator by using Eq. \eqref{hh}.
The detailed calculations of the partial derivative on Eq. \eqref{coss} can be found in the Appendix \ref{App2}.

Since the partial derivative on Eq. \eqref{coss} contains the $\Theta(\cdot)$ function (see Eq. \eqref{cos}), for $s_1<s_2$, the first part of the velocity correlation function can be divided into two parts to simplify the calculations, i.e.,
\begin{equation}\label{vv1}
\begin{split}
&\langle v(s_1)v(s_2)\rangle_1\\
&=\frac{2f_0^2}{\omega^2}e^{-\gamma s_1}e^{-\gamma s_2}\int_0^{s_1} \int_0^{s'_1} e^{\gamma s'_1}e^{\gamma s'_2}\\
&~~~\times\frac{d^2}{ds'_1ds'_2}\langle \cos(\omega t(s'_1))\cos(\omega(t(s'_2)))\rangle ds'_2ds'_1\\
&~~~+\frac{f_0^2}{\omega^2}e^{-\gamma s_1}e^{-\gamma s_2}\int_0^{s_1} \int_{s_1}^{s_2} e^{\gamma s'_1}e^{\gamma s'_2}\\
                             &~~~\times\frac{d^2}{ds'_1ds'_2}\langle \cos(\omega t(s'_1))\cos(\omega(t(s'_2)))\rangle ds'_2ds'_1.
\end{split}
\end{equation}
Inserting Eq. \eqref{coss} into the form Eq. \eqref{vv1},  one can get the expression of the first part of velocity correlation function in operation time $s$ (see Eq. \eqref{vv} in Appendix \ref{App2}).
Then by virtue of the subordination method Eqs. \eqref{hh} and \eqref{two} and the velocity correlation function in operation time in Eq. \eqref{vv}, the first part of velocity correlation function in Laplace space with small $\lambda_{1}$ and $\lambda_{2}$ is

\begin{equation}\label{sd}
\begin{split}
\langle &v(\lambda_1)v(\lambda_2)\rangle_1
\simeq C_0\frac{\Phi(\lambda_1)+\Phi(\lambda_2)-\Phi(\lambda_1+\lambda_2)}{\lambda_1\lambda_2\Phi(\lambda_1+\lambda_2)},
\end{split}
\end{equation}
where
\begin{equation}\label{c0}
\begin{split}
  C_0=\textrm{Re}\left[\frac{f_0^2\Phi(i\omega)}{2\omega^2(\gamma+\Phi(i\omega))}\right].
\end{split}
\end{equation}
Here $\textrm{Re}[\cdot]$ denotes the real part of a complex number.

After the inverse Laplace transform, the first part of velocity correlation function for large $t_1$ and  $t_2$ ($t_1<t_2$) is
\begin{equation}\label{2}
\begin{split}
&\langle v(t_1)v(t_2)\rangle_1\\
&\simeq \left\{
\begin{array}{ll}
  C_1\frac{\sin(\pi\alpha)}{\pi}B\left(\frac{t_1}{t_2};\alpha,1-\alpha \right),  & ~0<\alpha<1,  \\[5pt]
  C_2\tau_0^{\alpha-1}\left((t_2-t_1)^{1-\alpha}-t_2^{1-\alpha}\right), &~ 1<\alpha<2,
\end{array}\right.
\end{split}
\end{equation}
where $$C_1=\frac{f^2_0\omega^{\alpha-2}(\gamma\cos(\alpha\pi/2)+\omega^\alpha)}{2(\gamma^2+2\gamma\omega^\alpha\cos(\alpha\pi/2)+\omega^{2\alpha})}$$
and
$$C_2=\frac{f^2_0b_1}
{2\omega^2(\gamma^2+b_2)}$$
with $b_1=\tau^2_0\omega^2/{(\alpha-1)^2}+\tau^{2\alpha}_0|\Gamma(1-\alpha)|^2\omega^{2\alpha}
-\tau^\alpha_0|\Gamma(1-\alpha)|\omega^\alpha(\gamma\cos(\alpha\pi/2)+2\tau_0\omega/(\alpha-1)\sin(\alpha\pi/2))$,
$b_2=\tau^2_0\omega^2/{(\alpha-1)^2}+\tau^{2\alpha}_0|\Gamma(1-\alpha)|^2\omega^{2\alpha}
-2\tau^\alpha_0|\Gamma(1-\alpha)|\omega^\alpha(\gamma\cos(\alpha\pi/2)+\tau_0\omega/(\alpha-1)\sin(\alpha\pi/2))$.
The two parts of the velocity correlation function in Eqs. \eqref{1} and \eqref{2}, respectively, present the same expressions except for the coefficients. 
When the time-dependent periodic force $F(t)$ indeed brings a bias to the Langevin system (see Eq. \eqref{x-mean}), it is interesting to find that the velocity correlation function only increases with a fixed proportion. Therefore, as the free L\'{e}vy walk, the velocity correlation function decays at the power-law rate $t_2^{-\alpha}$ whenever $0<\alpha<1$ or $1<\alpha<2$ for a fixed $t_1$.

From the above discussions, one can note that the correlation structure of the velocity process is determined by the free Langevin system, and it remains unchanged for a time-dependent periodic force. The amplitude $f_0$ and frequency $\omega$ of the periodic force only impact the proportion of enlarging the velocity correlation function.
This amplification effect will also extend to the position correlation function and the MSDs.
In the next section, we will show the expressions of EAMSD and TAMSD of the L\'{e}vy-walk-like Langevin dynamics affected by the time-dependent periodic force and show its `ultraweak' non-ergodic behavior.

\section{EAMSD and TAMSD}\label{six}
Combining Eqs. \eqref{1} and \eqref{2}, we obtain the velocity correlation function with $t_1<t_2$ as
\begin{equation} \label{v-correlation}
\begin{split}
&\langle v(t_1)v(t_2)\rangle\\
&\simeq \left\{
\begin{array}{ll}
  \left(C_1+\frac{D}{\gamma}\right)\frac{\sin(\pi\alpha)}{\pi}B\left(\frac{t_1}{t_2};\alpha,1-\alpha \right),  & ~0<\alpha<1,  \\[5pt]
  \left(C_2+\frac{D}{\gamma}\right)\tau_0^{\alpha-1}\left((t_2-t_1)^{1-\alpha}-t_2^{1-\alpha}\right), &~ 1<\alpha<2.
\end{array}\right.
\end{split}
\end{equation}
Inserting  Eq. \eqref{v-correlation} into the equality
\begin{equation}\label{xx}
\langle x^2(t)\rangle=2\int_0^{t}\int_0^{t_2}\langle v(t_1)v(t_2)\rangle dt_1dt_2,
\end{equation}
one arrives at the second moment of the L\'{e}vy walk in the external periodic force field for large time $t$, i.e.,
\begin{equation}\label{EAMSD}
\langle x^2(t)\rangle\simeq \left\{
\begin{array}{ll}
  \left(C_1+\frac{D}{\gamma}\right)(1-\alpha)t^2,  & ~0<\alpha<1,  \\[5pt]
  2\left(C_2+\frac{D}{\gamma}\right)\frac{(\alpha-1)\tau_0^{\alpha-1}}{(2-\alpha)(3-\alpha)}t^{3-\alpha}, &~ 1<\alpha<2,
\end{array}\right.
\end{equation}
which is also the asymptotic expression of EAMSD, since the bias coming from the square of first moment $\langle x(t)\rangle^2$ in Eq. \eqref{x-mean} is far less than the second moments $\langle x^2(t)\rangle$. This EAMSD shows the ballistic diffusion when $0<\alpha<1$ and sub-ballistic superdiffusion when $1<\alpha<2$, the same as free L\'{e}vy walk;
but the external time-dependent periodic force $F(t)$ contributes to an additional dispersion on the particles' position here.

\begin{figure*}[!htb]
\begin{minipage}{0.2\linewidth}
  \centerline{\includegraphics[scale=0.295]{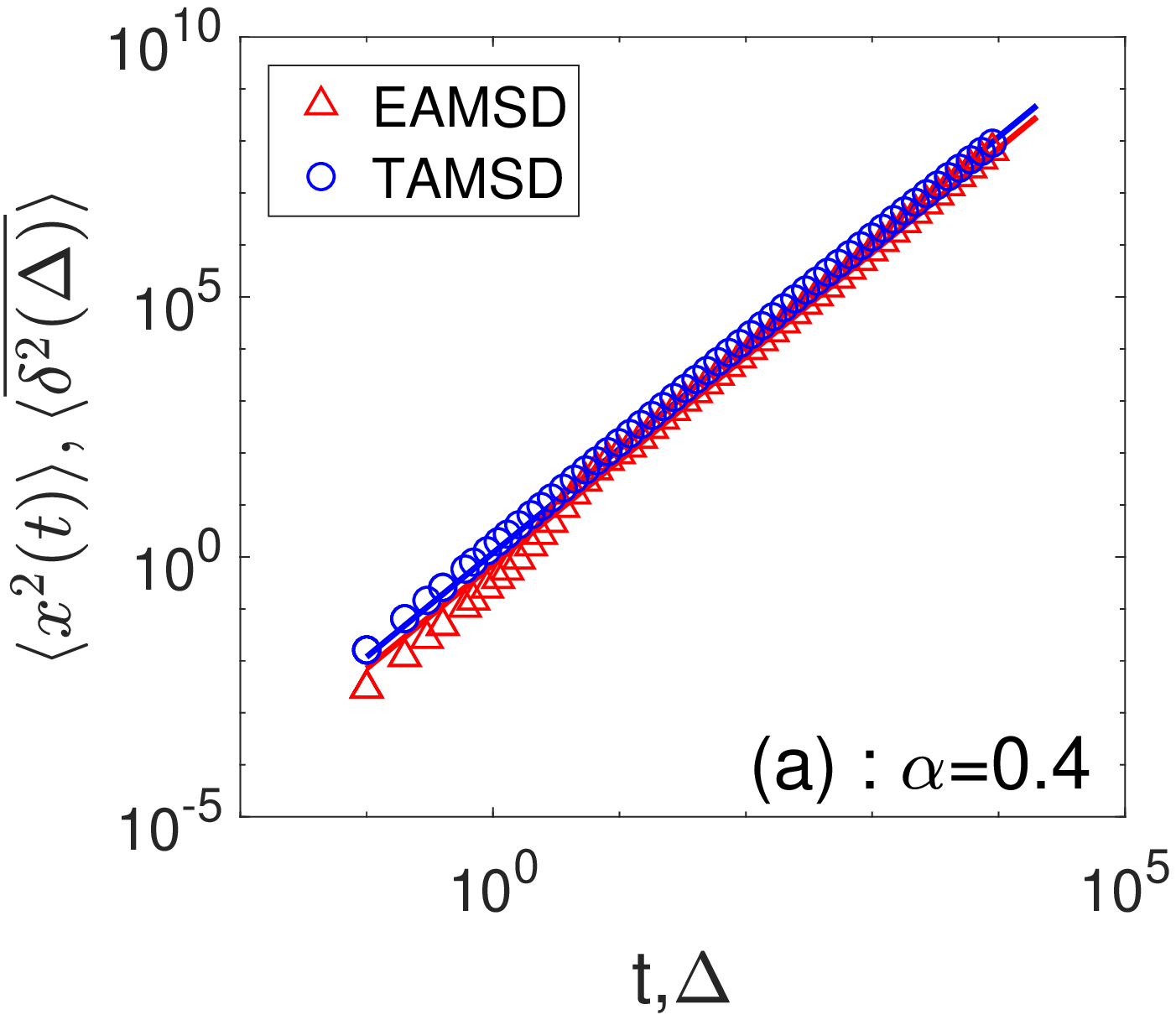}}
\end{minipage}
\hspace{6mm}
\begin{minipage}{0.2\linewidth}
  \centerline{\includegraphics[scale=0.295]{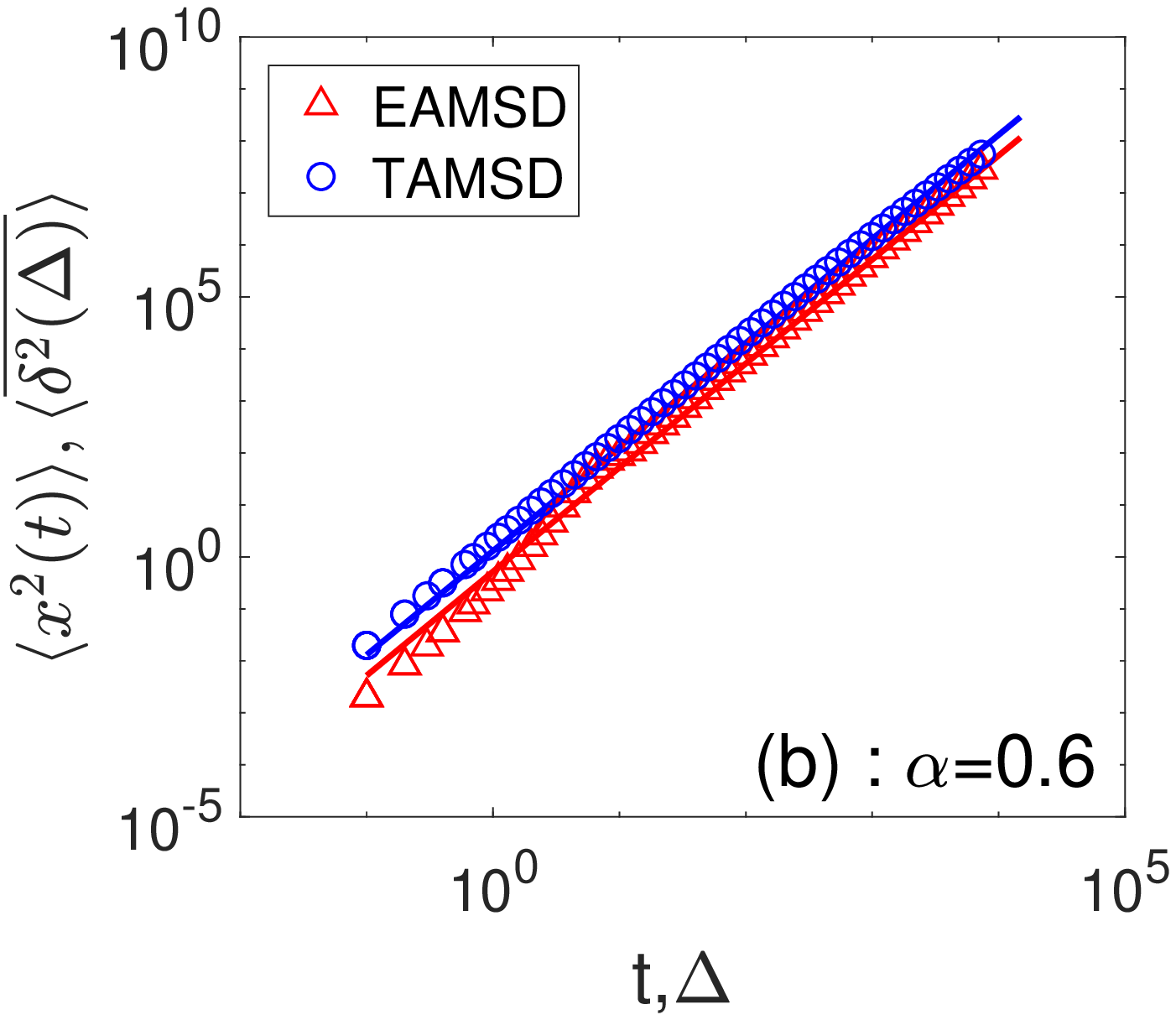}}
\end{minipage}
\hspace{6mm}
\begin{minipage}{0.2\linewidth}
  \centerline{\includegraphics[scale=0.295]{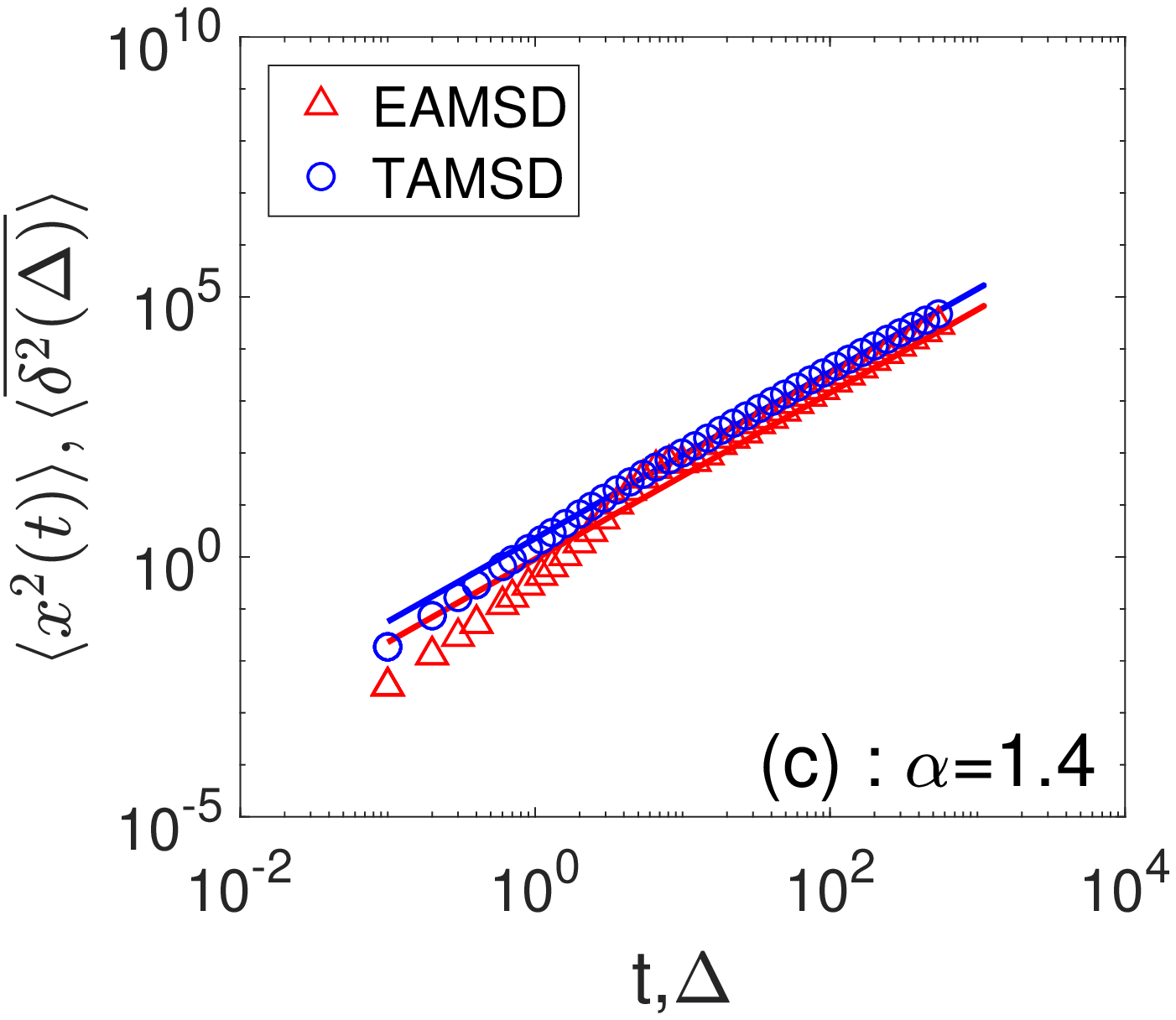}}
\end{minipage}
\hspace{6mm}
\begin{minipage}{0.2\linewidth}
  \centerline{\includegraphics[scale=0.295]{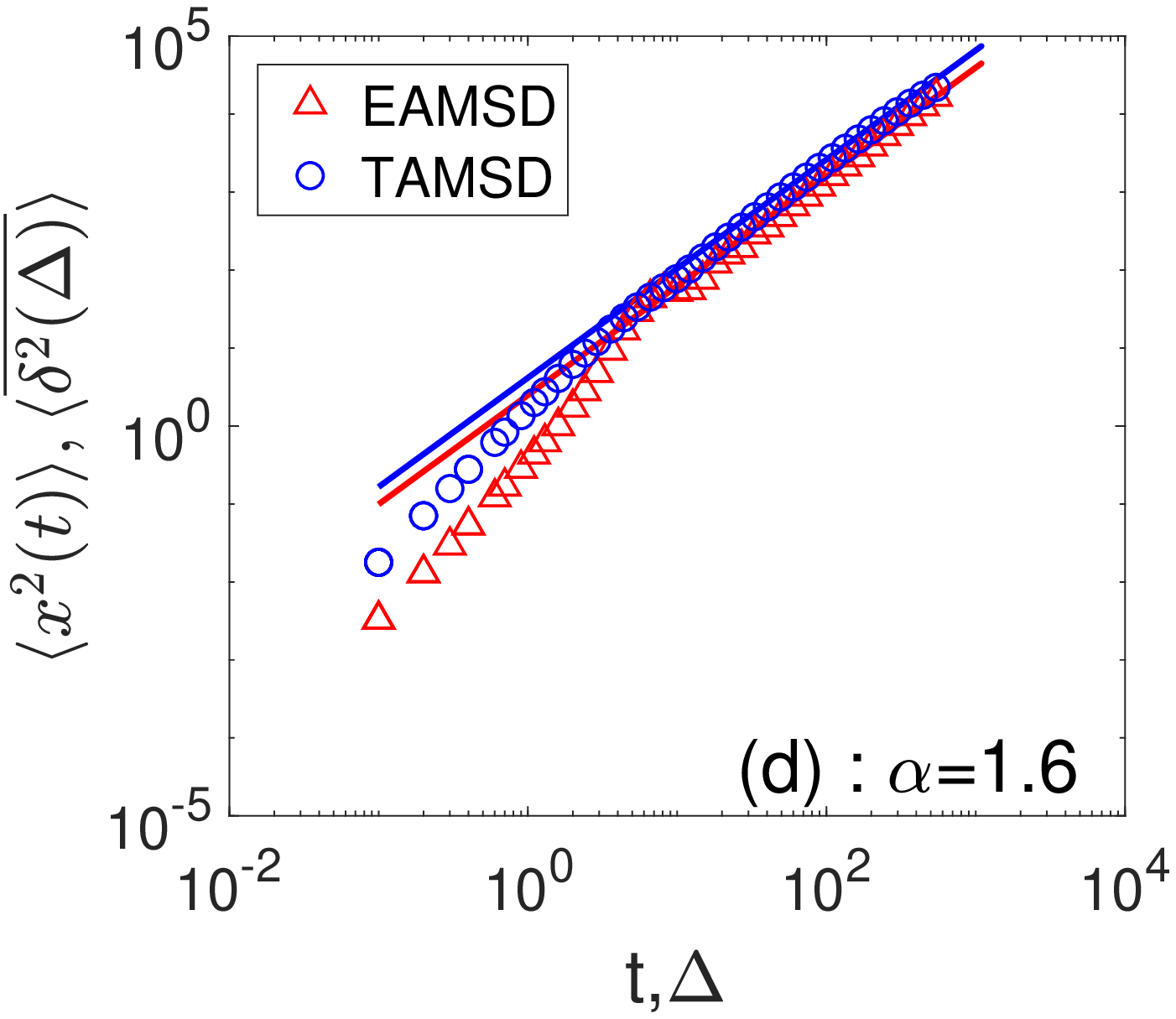}}
\end{minipage}
\caption{{(Color online)} Simulation results of the EAMSD and TAMSD for stochastic process described by the Langevin equation Eq. \eqref{LW_force} with different $\alpha$ and parameters $f_0=1$, $\omega=0.5$, $\gamma=2$, $\tau_0=1$.
Red triangle-markers and red solid lines are, respectively, the simulation and theoretical results Eq. \eqref{EAMSD} for EAMSD; the blue circle-markers and the blue solid lines are, respectively, the simulation and theoretical results Eq. \eqref{TAMSD} for TAMSD. }\label{fig}
\end{figure*}

As for the TAMSD for $\Delta\ll T$, the integrand $\langle (x(t+\Delta)-x(t))^2\rangle$ in the definition of TAMSD in Eq. \eqref{TADef} can be obtained by use of the velocity correlation function as $\int_t^{t+\Delta}\int_t^{t+\Delta} \langle v(t_1)v(t_2)\rangle dt_1dt_2$. Similar to the EAMSD, the bias coming from the first moment $\langle x(t)\rangle^2$ is far less than the second moments, and can be neglected. Then we obtain a result similar to the free L\'{e}vy walk:
\begin{equation}\label{TAMSD}
\langle \overline{\delta^2(\Delta)}\rangle\simeq \left\{
\begin{array}{ll}
  \left(C_1+\frac{D}{\gamma}\right)\Delta^2,  & ~0<\alpha<1,  \\[5pt]
  2\left(C_2+\frac{D}{\gamma}\right)\frac{\tau_0^{\alpha-1}}{(2-\alpha)(3-\alpha)}\Delta^{3-\alpha}, &~ 1<\alpha<2.
\end{array}\right.
\end{equation}
We show the simulation results of EAMSD and TAMSD in Fig. \ref{fig}, which coincide with the theoretical results in Eqs. \eqref{EAMSD} and \eqref{TAMSD} well. Equation \eqref{TAMSD} can also be obtained from the generalized Green-Kubo formula \cite{DechantLutzKesslerBarkai:2014,MeyerBarkaiKantz:2017}. The response of the L\'{e}vy-walk-like Langevin system to the time-dependent periodic force is similar to the subdiffusion case in CTRW model, where the force contribution makes an additional dispersion of the particle position compared with the force-free case \cite{SokolovKlafter:2006,MagdziarzWeronKlafter:2008,ChenWangDeng:2019-2}. The coefficients $C_{1,2}$ are contributed from the time-dependent periodic force $F(t)$. Taking the amplitude of the periodic force as $f_0=0$, the constant $C_{1,2}$ becomes zero and the statistical qualities above all reduce to the ones in free L\'{e}vy walk \cite{WangChenDeng:2019,FroembergBarkai:2013}.  In addition, the `ultraweak' non-ergodic behavior \cite{GodecMetzler:2013} is also observed here like the case of free L\'{e}vy walk, which means the TAMSD differs EAMSD only by a constant.


\section{Generalized Klein-Kramers equation}\label{seven}

In the above sections, we show how the L\'{e}vy-walk-like Langevin dynamics is affected by an external time-dependent periodic force by evaluating various statistical quantities. In this section, we derive the generalized Klein-Kramers equation satisfied by the joint PDF $P(x,v,t)$  of finding the particle at position $x$ with velocity $v$ at time $t$ by using It\^{o} formula, which has been used to derive the Feynman-Kac equation \cite{CairoliBaule:2017}. We take the time-dependent force as an example; the detailed derivations and results can also be applied to a more general force, such as a constant force or a linear force.

Thanks to the finite variation of $t(s)$, both $v(s)$ and the time-changed process $v(t):=v(s(t))$ are semi-martingales \cite{CairoliBaule:2017}. Then the It\^{o} formula of a semi-martingale $w(t)=(v(t), x(t))$, as 
\begin{equation}\label{ito}
\begin{split}
&f(w(t))
=f(w_0)+\int_0^t \frac{\partial}{\partial v}f(w(\tau))d v(\tau)\\
&~+\int_0^t \frac{\partial}{\partial x}f(w(\tau))d x(\tau)
+\frac{1}{2}\int_0^t \frac{\partial^2}{\partial v\partial x}f(w(\tau))d [v,x]_\tau\\
&~+\frac{1}{2}\int_0^t \frac{\partial^2}{\partial v^2}f(w(\tau))d [v,v]_\tau
+\frac{1}{2}\int_0^t \frac{\partial^2}{\partial x^2}f(w(\tau))d [x,x]_\tau, 
\end{split}
\end{equation}
can be used to derive the generalized Klein-Kramers equation. The covariation $[v,x]_t$ and quadratic variation $[x,x]_t$ are both null, since $x(t)=\int_0^t v(\tau)d\tau$ is a finite variance process. In addition, the quadratic variation of the time-changed velocity process $v(t)$ is $[v,v]_t=s(t)$ according to Eq. \eqref{dvt}. Taking $f(w(t))=e^{ikv(t)+ipx(t)}$, by use of Eq. \eqref{dvt} of the velocity process $v(t)$ and the It\^{o} formula \eqref{ito}, one has
\begin{equation}
\begin{split}
&f(w(t))
=f(w_0)-ik\gamma \int_0^tf(w(\tau)) v(\tau)\dot{s}(\tau)d\tau\\
&+ik\int_0^tf(w(\tau)) F(\tau)d\tau
+ik\int_0^tf(w(\tau))\xi(s(\tau))\dot{s}(\tau)d\tau\\
&+ip\int_0^tf(w(\tau))v(\tau)d\tau
-\frac{k^2}{2}\int_0^tf(w(\tau))\dot{s}(\tau)d\tau.
\end{split}
\end{equation}
Taking the ensemble average over the realizations on both sides of the above equation, one has
\begin{equation}\label{i}
\begin{split}
\langle f(w(t))\rangle
&=\langle f(w_0)\rangle+ip\int_0^t\langle f(w(\tau))v(\tau)\rangle d\tau \\
&~~~+ik\int_0^t\langle f(w(\tau))\rangle F(\tau)d\tau\\
&~~~+\left\langle \int_0^t f(w(\tau))\left[- ik \gamma v(\tau)-\frac{1}{2}k^2\right]\dot{s}(\tau)d\tau\right\rangle,
\end{split}
\end{equation}
since for each fixed realization of $s$, the term
$\langle\int_0^tf(w(\tau))\xi(s(\tau))\dot{s}(\tau)d\tau\rangle$
is null due to the zero mean and independence of the increments of Brownian motion. The left side of Eq. \eqref{i} is exactly the Fourier transform of PDF $P(x,v,t)$, i.e., $P(p,k,t)$.
Then performing the inverse Fourier transform $(k\rightarrow v)$ towards Eq. \eqref{i} and taking partial derivative with respect to time $t$, one arrives at
\begin{equation}\label{ii}
\begin{split}
&\frac{\partial}{\partial t}P(p,v,t)\\
&=\left(ipv-\frac{\partial}{\partial v}F(t)\right)P(p,v,t)
+\left(\gamma \frac{\partial}{\partial v}v+\frac{1}{2}\frac{\partial^2}{\partial v^2}\right)\\
&~~~\times\frac{\partial}{\partial t}\left\langle \int_0^t e^{ipx(\tau)}\delta(v-v(\tau))\dot{s}(\tau)d\tau\right\rangle.\end{split}
\end{equation}
The remaining difficulty of deriving the generalized Klein-Kramers equation is how to build the relation between the last term
\begin{equation*}
\frac{\partial}{\partial t}\left\langle \int_0^t e^{ipx(\tau)}\delta(v-v(\tau))\dot{s}(\tau)d\tau\right\rangle
 \end{equation*}
and ${P}(p,v,t)$, then the inverse Fourier transform $(p\rightarrow x)$ can be performed. By virtue of the important relation (see Appendix \ref{App3} for the detailed derivation)
\begin{equation}\label{relation}
\begin{split}
&\mathcal{L}_{t\rightarrow\lambda}\left[\frac{\partial}{\partial t}\left\langle \int_0^t e^{ipx(\tau)}\delta(v-v(\tau))\dot{s}(\tau)d\tau\right\rangle\right] \\
&~~~~~~~=\frac{\lambda-ipv}{\Phi(\lambda-ipv)}P(p,v,\lambda),
\end{split}
\end{equation}
we can finally obtain the generalized Kleins-Kramers equation corresponding to the L\'{e}vy-walk-like Langevin dynamics in Eq. \eqref{LW_force} after substituting the equality Eq. \eqref{relation} into Eq. \eqref{ii} and taking the inverse Fourier and Laplace  transforms $(p\rightarrow x, \lambda\rightarrow t)$:
\begin{equation}\label{KK}
\begin{split}
&\left(\frac{\partial}{\partial t}+\frac{\partial}{\partial x}v
+\frac{\partial}{\partial v}F(t)\right)P(x,v,t)\\
&~=\left(\gamma \frac{\partial}{\partial v}v+\frac{1}{2}\frac{\partial^2}{\partial v^2}\right)\left(\frac{\partial}{\partial t}+\frac{\partial}{\partial x}v\right) \\
&~~~~~\times\int_0^t K(t-\tau)P(x-v(t-\tau), v,\tau)d\tau.
\end{split}
\end{equation}
Here the Laplace transform of $K(t)$ is $K(\lambda)=1/\Phi(\lambda)$.
With the inverse Laplace transform, one has $K(t)=t^{\alpha-1}/\Gamma(\alpha)$ for $0<\alpha<1$ and $K(t)=(\alpha-1)/\tau_0+\tau^{\alpha-2}_0(\alpha-1)t^{1-\alpha}$ for $1<\alpha<2$.

Taking the external force $F(t)=0$, the generalized Kramers-Fokker-Planck equation for CTRW in position-velocity space is recovered \cite{FriedrichJenkoBauleEule:2006,FriedrichJenkoBauleEule:2006-2}.
Note that our derivation of the generalized Kleins-Kramers equation is also valid for other kind of external force, only if this force is acting on the system for all physical time $t$. For the constant force $F(x)\equiv f_0$ or the harmonic potential with force being $F(x)=-x(t(s))$, the corresponding generalized Klein-Kramers equation can be obtained by replacing $F(t)$ in Eq. \eqref{KK} with the new force $F(x)$.
In addition, we find that the memory kernel $K(t)$ in the integral of the generalized Klein-Kramers equation Eq. \eqref{KK} comes from the power-law distributed flight time
of L\'{e}vy walk, and it is independent with the external forces. This phenomenon is different from the subdiffusion case in CTRW model, where the external forces influence not only the drift term, but also the integral operator \cite{CairoliKlagesBaule:2018,ChenWangDeng:2019-2}.

\section{Summary}\label{eight}
L\'{e}vy walk is originally proposed as a coupled CTRW model and then much related research work has been undertaken. Langevin picture is an alternative way to describe the L\'{e}vy-walk-like dynamics. The significant advantage of Langevin picture is the convenience of including an external force, evaluating the correlation function, and modeling the time changed process. The situation of L\'{e}vy walk under a space-dependent force or a constant force has been discussed before \cite{BarkaiFleurov:1998,ChenWangDeng:2019-3,WangChenDeng:2020PRE,XuZhouMetzlerDeng:2020}. Here, this paper aims at investigating the response of L\'{e}vy-walk-like Langevin dynamics to an external time-dependent periodic force.



Although the external force is periodic, the first moment of the particle displacement $\langle x(t)\rangle$ is not longer zero even for a long time. Compared with the constant force or harmonic potential acting on L\'{e}vy walk, where the diffusion behavior is significantly changed, the time-dependent periodic force looks more mild. It only increases the (generalized) diffusion coefficient, but remains the diffusion structure.
This phenomenon has some similarity to the case in which the time-dependent periodic force acts on the subdiffusive CTRW over operation time \cite{SokolovKlafter:2006,ChenWangDeng:2019-2}. Furthermore, the weak difference between EAMSD and TAMSD of L\'{e}vy walk under time-dependent periodic force is retained here, and
the `ultraweak' non-ergodic behavior of free L\'{e}vy walk also exists. The results of the statistical quantities in this paper can recover the ones of the force-free case by taking the external force to be zero.

Based on the It\^{o} formula, we derive the corresponding generalized Klein-Kramers equation including a time-dependent force. Replacing $F(t)$ with a general force term $F(x)$, the constant force or the linear force for harmonic potential, the generalized Klein-Kramers equation is also valid. The memory kernel in the generalized Klein-Kramers equation comes from the power-law distributed flight time of L\'{e}vy walk, and it does not interact with the arbitrary external force $F(x,t)$, which is also a significance difference from the subdiffusive CTRW \cite{CairoliKlagesBaule:2018,ChenWangDeng:2019-2}.

\section*{Acknowledgments}
This work was supported by the National Natural Science Foundation of China under grant no. 12071195, and the Fundamental Research Funds for the Central Universities under grant no. lzujbky-2020-it02.

\appendix
\begin{widetext}
\section{Derivations of Eqs. \eqref{dvt} and \eqref{vt}}\label{App1}
Let us firstly give the detailed derivation of Eq. \eqref{dvt}. Replacing $s$ in the second equation in Eq. \eqref{LW_force} with $s(t)$, one gets
\begin{equation}
\begin{split}
\frac{d}{d s(t)}v(s(t))=-\gamma v(s(t))+F(t) \eta(s(t))+\xi(s(t)),
\end{split}
\end{equation}
where we have used the fact that $t(s(t))\equiv t$. Then using the relation $v(t):=v(s(t))$ and multiplying $ds(t)/dt$ on both sides of the above equation, one has
\begin{equation}\label{a}
\begin{split}
\frac{d}{dt}v(t)&=-\gamma v(s(t))\frac{d}{dt}s(t)+F(t) \eta(s(t))\frac{d}{dt}s(t)+\xi(s(t))\frac{d}{dt}s(t)\\
                &=-\gamma v(t)\frac{d}{dt}s(t)+F(t)+\xi(s(t))\frac{d}{dt}s(t),
\end{split}
\end{equation}
where the relationship
\begin{equation}
  \eta(s(t))=\frac{dt(s(t))}{ds(t)}= \frac{dt}{ds(t)}
\end{equation}
has been used in the last line in Eq. \eqref{a}.

As for Eq. \eqref{vt}, replacing $s$  with $s(t)$ in Eq. \eqref{vs} gives
\begin{equation}
\begin{split}
v(t):&=v(s(t))\\
     &=\int_0^{s(t)} e^{-\gamma(s(t)-s')}F(t(s'))\eta(s')ds'+\int_0^{s(t)} e^{-\gamma(s(t)-s')}\xi(s')ds'\\
     &=\int_0^t e^{-\gamma(s(t)-s(t'))}F(t')\eta(s(t'))ds(t')+\int_0^t e^{-\gamma(s(t)-s(t'))}\xi(s(t'))ds(t')\\
     &=\int_0^t e^{-\gamma(s(t)-s(t'))}F(t')dt'+\int_0^t e^{-\gamma(s(t)-s(t'))}\xi(s(t'))ds(t'),
\end{split}
\end{equation}
where we perform the variable substitution $s'\rightarrow s(t')$ in the third line.

\section{Velocity correlation
function $\langle v(s_1)v(s_2)\rangle_1$}\label{App2}

To obtain the velocity correlation
function, we firstly present the specific expression of Eq. \eqref{coss}:
\begin{equation}\label{cos}
\begin{split}
&\frac{d^2}{ds_1ds_2}\langle \cos(\omega t(s_1))\cos(\omega t(s_2))\rangle\\
&=\frac{1}{4}\frac{d^2}{ds_1ds_2}[p(-i\omega,s_1;-i\omega,s_2)+p(-i\omega,s_1;i\omega,s_2)
+p(i\omega,s_1;-i\omega,s_2)+p(i\omega,s_1;i\omega,s_2)]\\
&=\frac{1}{4}\delta(s_2-s_1)\left[(\Phi(-i\omega)-\Phi(-2i\omega)+\Phi(-i\omega))e^{-\Phi(-2i\omega)s_1}
+2(\Phi(-i\omega)+\Phi(i\omega))\right.\\
&~~~~~~~~~~~~~~~~~~~~+\left.(\Phi(i\omega)-\Phi(2i\omega)+\Phi(i\omega))e^{-\Phi(2i\omega)s_1}\right]\\
&~~~+\frac{1}{4}\Theta(s_2-s_1)\left[\Phi(-i\omega)(\Phi(-2i\omega)-\Phi(-i\omega))e^{-\Phi(-2i\omega)s_1}e^{-\Phi(-i\omega)(s_2-s_1)}\right.\\
&~~~~~~~~~~~~~~~~~~~~~~~-\Phi^2(i\omega)e^{-\Phi(i\omega)(s_2-s_1)}
-\Phi^2(-i\omega)e^{-\Phi(-i\omega)(s_2-s_1)}\\
&~~~~~~~~~~~~~~~~~~~~~~~+\left.\Phi(i\omega)(\Phi(2i\omega)-\Phi(i\omega))e^{-\Phi(2i\omega)s_1}e^{-\Phi(i\omega)(s_2-s_1)}\right]\\
&~~~+\frac{1}{4}\Theta(s_1-s_2)\left[\Phi(-i\omega)(\Phi(-2i\omega)-\Phi(-i\omega))e^{-\Phi(-2i\omega)s_2}e^{-\Phi(-i\omega)(s_1-s_2)}\right.\\
&~~~~~~~~~~~~~~~~~~~~~~~-\Phi^2(i\omega)e^{-\Phi(i\omega)(s_1-s_2)}
-\Phi^2(-i\omega)e^{-\Phi(-i\omega)(s_1-s_2)}\\
&~~~~~~~~~~~~~~~~~~~~~~~+\left.\Phi(i\omega)(\Phi(2i\omega)-\Phi(i\omega))e^{-\Phi(2i\omega)s_2}e^{-\Phi(i\omega)(s_1-s_2)}\right].
\end{split}
\end{equation}
Inserting Eq. \eqref{cos} into Eq. \eqref{vv1}, one has the lengthy expression of the first part of velocity correlation function in operation time $s$:
\begin{equation}\label{vv}
\begin{split}
\langle v(s_1)v(s_2)\rangle_1
=&O_-\cdot\left[e^{(\gamma-\Phi(-2i\omega))s_1}e^{-\gamma s_2}-e^{-\gamma(s_1+s_2)}\right]+P\cdot\left[e^{-\gamma(s_2-s_1)}-e^{-\gamma(s_1+s_2)}\right]\\
                              &+ O_+\cdot\left[e^{(\gamma-\Phi(2i\omega))s_1}e^{-\gamma s_2}-e^{-\gamma(s_1+s_2)}\right]\\
                              &+Q_-\cdot\left[\frac{e^{(\gamma-\Phi(-2i\omega))s_1}e^{-\gamma s_2}-e^{-\gamma(s_1+s_2)}}{2\gamma-\Phi(-2i\omega)}-\frac{e^{-\Phi(-i\omega)s_1}e^{-\gamma s_2}-e^{-\gamma(s_1+s_2)}}{\gamma-\Phi(-i\omega)}\right]\\
                              &-R_-\cdot\left[\frac{e^{-\gamma(s_2-s_1)}-e^{-\gamma(s_1+s_2)}}{2\gamma}-\frac{e^{-\Phi(-i\omega)s_1}e^{-\gamma s_2}-e^{-\gamma(s_1+s_2)}}{\gamma-\Phi(-i\omega)}\right]\\
                              &-R_+\cdot\left[\frac{e^{-\gamma(s_2-s_1)}-e^{-\gamma(s_1+s_2)}}{2\gamma}-\frac{e^{-\Phi(i\omega)s_1}e^{-\gamma s_2}-e^{-\gamma(s_1+s_2)}}{\gamma-\Phi(i\omega)}\right]\\
                              &+Q_+\cdot\left[\frac{e^{(\gamma-\Phi(2i\omega))s_1}e^{-\gamma s_2}-e^{-\gamma(s_1+s_2)}}{2\gamma-\Phi(2i\omega)}-\frac{e^{-\Phi(i\omega)s_1}e^{-\gamma s_2}-e^{-\gamma(s_1+s_2)}}{\gamma-\Phi(i\omega)}\right]\\
                              &+S_-\cdot\left[e^{(\Phi(-i\omega)-\Phi(-2i\omega))s_1}e^{-\Phi(-i\omega)s_2}-e^{-\gamma s_1}e^{-\Phi(-i\omega)s_2}\right.\\
                              &~~~~~~~~~\left.-e^{(\gamma-\Phi(-2i\omega))s_1}e^{-\gamma s_2}+e^{-\Phi(-i\omega)s_1}e^{-\gamma s_2}\right]\\
                              &-T_+\cdot\left[e^{\Phi(i\omega)s_1}e^{-\Phi(i\omega)s_2}-e^{-\gamma s_1}e^{-\Phi(i\omega)s_2}-e^{\gamma s_1}e^{-\gamma s_2}+e^{-\Phi(i\omega)s_1}e^{-\gamma s_2}\right]\\
                              &-T_-\cdot\left[e^{\Phi(-i\omega)s_1}e^{-\Phi(-i\omega)s_2}-e^{-\gamma s_1}e^{-\Phi(-i\omega)s_2}-e^{\gamma s_1}e^{-\gamma s_2}+e^{-\Phi(-i\omega)s_1}e^{-\gamma s_2}\right]\\
                              &+S_+\cdot\left[e^{(\Phi(i\omega)-\Phi(2i\omega))s_1}e^{-\Phi(i\omega)s_2}-e^{-\gamma s_1}e^{-\Phi(i\omega)s_2}\right.\\
                              &~~~~~~~~~\left.-e^{(\gamma-\Phi(2i\omega))s_1}e^{-\gamma s_2}+e^{-\Phi(i\omega)s_1}e^{-\gamma s_2}\right],
\end{split}
\end{equation}
where
\begin{equation}
\begin{split}
O_\pm&=\frac{f_0^2}{4\omega^2}\frac{\Phi(\pm i\omega)-\Phi(\pm2i\omega)+\Phi(\pm i\omega)}{2\gamma -\Phi(\pm2i\omega)},\\
P&=\frac{f_0^2}{4\omega^2}\frac{\Phi(i\omega)+\Phi(-i\omega)}{\gamma},\\
Q_\pm&=\frac{f_0^2}{2\omega^2}\frac{\Phi(\pm i\omega)(\Phi(\pm2i\omega)-\Phi(\pm i\omega))}{\gamma-\Phi(\pm2i\omega)+\Phi(\pm i\omega)},\\
R_\pm&=\frac{f_0^2}{2\omega^2}\frac{\Phi^2(\pm i\omega)}{\gamma +\Phi(\pm i\omega)},\\
S_\pm&=\frac{f_0^2}{4\omega^2}\frac{\Phi(\pm i\omega)(\Phi(\pm 2i\omega)-\Phi(\pm i\omega))}{(\gamma-\Phi(\pm i\omega))(\gamma+\Phi(\pm i\omega)-\Phi(\pm 2i\omega))},\\
T_\pm&=\frac{1}{2(\gamma-\Phi(\pm i\omega))}R_\pm.
\end{split}
\end{equation}
To obtain the first part of velocity correlation function in Laplace space, we will apply the subordination formulae in Eq. \eqref{two}.
Although the expression of Eq. \eqref{vv} looks very complicated, it can be greatly simplified for long time limit. Since the long time corresponds to large $s_1$ and $s_2$, most of the terms in Eq. \eqref{vv} decay exponentially, except for those containing $e^{-c(s_2-s_1)}$ ($c=\gamma$ or $\Phi(\pm i\omega)$). Therefore, we remain these terms, multiply them with the two-point PDF of inverse subordinator in Eq. \eqref{hh}, and perform the integral. Furthermore, considering the exponential kernel $e^{-c(s_2-s_1)}$ in integral, the dominant term from Eq. \eqref{hh} is the first term containing $\delta(s_1-s_2)$, which will also simplify the calculations. In fact, we found the latter technique and applied it to other models in Ref. \cite{ChenWangDeng:2019-3}. Finally, for small $\lambda_1$ and $\lambda_2$, one obtains
\begin{equation}\label{aaaa}
\begin{split}
\langle v(\lambda_1)v(\lambda_2)\rangle_1
&=\int_0^\infty \int_0^\infty\langle v(s_1)v(s_2)\rangle h(s_1,\lambda_1;s_2,\lambda_2)ds_1 ds_2\\
&\simeq \left[P-\frac{R_-+R_+}{2\gamma}\right]\frac{\Phi(\lambda_1)+\Phi(\lambda_2)-\Phi(\lambda_1+\lambda_2)}{\lambda_1\lambda_2\Phi(\lambda_1+\lambda_2)},
\end{split}
\end{equation}
where the coefficient $P-\frac{R_-+R_+}{2\gamma}$ can be rewritten as $C_0:=\textrm{Re}\left[\frac{f_0^2\Phi(i\omega)}{2\omega^2(\gamma+\Phi(i\omega))}\right]$ in Eq. \eqref{c0}.


\section{Derivation of the relationship Eq. \eqref{relation}}\label{App3}
We use the method in Ref. \cite{CairoliBaule:2017} to obtain the relationship Eq. \eqref{relation}, i.e.,
\begin{equation}
\begin{split}
\mathcal{L}_{t\rightarrow\lambda}\left[\frac{\partial}{\partial t}\left\langle \int_0^t e^{ipx(\tau)}\delta(v-v(\tau))\dot{s}(\tau)d\tau\right\rangle\right]
=\frac{\lambda-ipv}{\Phi(\lambda-ipv)}P(p,v,\lambda).
\end{split}
\end{equation}
Firstly, let us rewrite the stochastic process $x(t)$ in physical time to the form
\begin{equation}\label{xA}
x(t)=A(s(t)),
\end{equation}
where $A(s)$ can be regarded as the original process of $x(t)$ in operation time $s$ with the expression $A(s)=\int_0^s v(\tau)\eta(\tau)d\tau$. Equation \eqref{xA} can be verified as
\begin{equation}
\begin{split}
A(s(t))=\int_0^{s(t)} v(\tau)\eta(\tau)d\tau=\int_0^{t} v(s(t'))\eta(s(t'))ds(t')
=\int_0^{t} v(t')dt'=x(t).
\end{split}
\end{equation}
Then, applying the equality $\delta(s-s(\tau))\dot{s}(\tau)=\delta(\tau-t(s))$, one can make the following arrangement:
\begin{equation}
\begin{split}
\int_0^t e^{ipx(\tau)}\delta(v-v(\tau))\dot{s}(\tau)d\tau
&=\int_0^t \left[\int_0^\infty e^{ipA(s)}\delta(v-v(s))\delta(s-s(\tau))ds\right]\dot{s}(\tau)d\tau\\
&=\int_0^t \left[\int_0^\infty e^{ipA(s)}\delta(v-v(s))\delta(\tau-t(s))ds\right]d\tau.
\end{split}
\end{equation}
Performing the ensemble average on both sides and taking the partial derivative with respect to $t$, we obtain
\begin{equation}\label{555}
\begin{split}
\frac{\partial}{\partial t}\left\langle\int_0^t e^{ipx(\tau)}\delta(v-v(\tau))\dot{s}(\tau)d\tau\right\rangle
&=\int_0^\infty \left \langle e^{ipA(s)}\delta(v-v(s))\delta(t-t(s))\right\rangle  ds \\
&=\mathcal{L}_{\lambda\rightarrow t}^{-1} \left[ \int_0^\infty \left \langle e^{-\lambda t(s)+ipA(s)}\delta(v-v(s))\right\rangle ds \right].
\end{split}
\end{equation}

On the other hand, applying Eq. \eqref{xA} again, we can rewrite the definition of $P(p,v,t)$ in the following form
\begin{equation}\label{dys}
\begin{split}
{P}(p,v,t)=\left\langle e^{ipA(s(t))} \delta(v-v(s(t)))\right\rangle
=\int_0^\infty \left\langle e^{ipA(s)}\delta(v-v(s))\delta(s-s(t))\right\rangle ds.
\end{split}
\end{equation}
Then we need to get the expression of the PDF $P(p,v,t)$ in Laplace space ($t\rightarrow\lambda$), i.e., ${P}(p,v,\lambda)$. For this, we firstly calculate the integral as
\begin{equation}
\begin{split}
\int_0^\infty \delta(s-s(t))e^{-\lambda t}dt
&=\frac{\partial}{\partial s}\int_0^\infty \Theta(s-s(t))e^{-\lambda t}dt  \\
&=\frac{\partial}{\partial s}\int_0^\infty [1-\Theta(t-t(s))]e^{-\lambda t}dt\\
&=\frac{\partial}{\partial s}\int_0^{t(s)}e^{-\lambda t}dt
=\eta(s)e^{-\lambda t(s)}.
\end{split}
\end{equation}
Then we obtain
\begin{equation}\label{dysL}
\begin{split}
{P}(p,v,\lambda)=
\int_0^\infty \left\langle \delta(v-v(s))\eta(s)e^{-\lambda t(s)+ipA(s)}\right\rangle ds.
\end{split}
\end{equation}
The symbol of ensemble average in Eq. \eqref{dysL} is working for noises $\eta$ and $\xi$, respectively. The independence of these two noises allows us to perform the ensemble average on any one of the noises firstly.
Calculating the ensemble average of noise $\eta$ in Eq. \eqref{dysL} leads to
\begin{equation}\label{pjz}
\begin{split}
\left\langle \eta(s)e^{-\lambda t(s)+ipA(s)}\right\rangle
&=\left\langle \eta(s)e^{-\int_0^s[\lambda -ip v(r)]\eta(r)dr}\right\rangle
=-\frac{1}{\lambda -ip v(s)}\frac{\partial}{\partial s} \left\langle e^{-\int_0^s[\lambda -ip v(r)]\eta(r)dr} \right\rangle\\
&=-\frac{1}{\lambda -ip v(s)}\frac{\partial}{\partial s} e^{-\int_0^s\Phi(\lambda -ip v(r))dr}
=\frac{\Phi(\lambda -ip v(s))}{\lambda -ip v(s)}\left\langle e^{-\int_0^s[\lambda -ip v(r)]\eta(r)dr}\right \rangle,
\end{split}
\end{equation}
where we have used the characteristic function of the subordinator in the third equal sign. Then, substituting Eq. \eqref{pjz} into Eq. \eqref{dysL}, we have
\begin{equation}
\begin{split}
P(p,v,\lambda)
=\frac{\Phi(\lambda -ip v)}{\lambda -ip v}\int_0^\infty \left\langle e^{-\lambda t(s)+ipA(s)}\delta(v-v(s))\right\rangle ds.
\end{split}
\end{equation}
The integral term in above equation is exactly the one in Eq. \eqref{555}. Therefore, performing the Laplace transform on both sides of Eq. \eqref{555} yields the important relation
\begin{equation}\label{impor}
\begin{split}
\mathcal{L}_{t\rightarrow\lambda}\left[\frac{\partial}{\partial t}\left\langle \int_0^t e^{ipx(\tau)}\delta(v-v(\tau))\dot{s}(\tau)d\tau\right\rangle\right]
=\frac{\lambda-ipv}{\Phi(\lambda-ipv)}P(p,v,\lambda).
\end{split}
\end{equation}

\end{widetext}

\section*{References}
\bibliographystyle{apsrev4-1}
\bibliography{ReferenceW}

\end{document}